\documentstyle[12pt,epsfig]{article}
\begin{document}

\title{The Layer Phase in the Non-isotropic
  SU(3) Gauge Model at Finite Temperature}
\author{Guo-Li Wang$^a$, Ying-Kai Fu$^b$}
\maketitle
\begin{center}
$a$, Department of Physics, FuJian Normal University,
 FuZhou 350007, China\\
$b$, Department of Physics, JiLin University,
 ChangChun 130023, China\\
\end{center}
{\tiny
\begin{abstract}
The phase structure of a non-isotropic non-Abelian SU(3) lattice gauge model at finite 
temperature is investigated to the third order in the variational-cumulant
 expansion (VCE) approach.
The layer phase exists in this model in the cases
 of dimensions $D=4$, $D=5$ $(d=D-1)$.
 \end{abstract}}
\indent
\section{Introduction}

Up to now , though all experiments do not deny that the universe people live in a 
(3+1)-dimensional space-time, we need not to believe absolutely that our universe must
be so\cite{th9407011}. Perhaps our universe is only a domain of a higher dimensional space-time, which we 
do not realize at this moment. This thought is inspired by the quickly development
 of string theory recently.  Consequently, there must be critical differences in 
physics between our (3+1)-dimensional universe inside and outside. New phenomena need 
to be investigated to some extent.

An interesting non-isotropic Abelian lattice gauge model on the existence of a 
high dimensional space-time has been proposed in Ref.\cite{fu} by one of us, Fu, and
Nielsen, where the coupling constants of gauge interactions depend on the spatial-time
directions.
A new phase, $d$-dimensional layer phase, has been found in a 
$D$-dimensional $(D>d)$ non-isotropic U(1) lattice gauge model at 
zero temperature by using mean field method and strong coupling corrections.
 The results have been confirmed by 
Monte-Carlo (MC) simulation\cite{berman} and by other method, e.g., by using the 
variational-cumulant expansion approach\cite{liang}. Further investigation show 
that the layer phase exists in the non-isotropic Z(2)\cite{xiu},
 non-Abelian SU(2)\cite{liang}
(confirmed by MC simulation\cite{berman}) and SU(3)\cite{fu3}
lattice gauge model, also exists in the non-isotropic U(1) lattice gauge model with
bosons\cite{man} and fermions\cite{alte,nico} at zero temperature. Furthermore,
recent works confirm the exists of layer phase in non-isotropic U(1) lattice gauge model
 in Ref\cite{la1} and show that the layer physe exists in the non-isotropic U(1) lattice 
gauge model with Higgs\cite{la2}.

The phase structure of non-isotropic U(1)\cite{lun} and SU(2)\cite{lun2}
 lattice gauge model at finite temperature
has been studied with the variational-cumulant expansion model,
the layer phase is also found which is different from the zero temperature case.
In this paper, we shall use the variational-cumulant expansion approach\cite{tan} to
investigate the phase structure of the non-isotropic non-Abelian SU(3) lattice gauge
model at finite temperature. 

\section{The non-isotropic SU(3) theory using the VCE at finite temperature}

If we suppose that there is a $D$-dimensional space-time described by SU(3) gauge
symmetry with the coupling constants depending on the spatial-time directions.
Then the Wilson action of this non-isotropic SU(3) lattice pure gauge model
 is: 
 \begin{equation}
 S=\sum\limits_{\mu\nu}\frac{\beta_{\mu\nu}}{2N}\sum\limits_{\{p_{\sigma}\mu\nu\}}
 ReTr(U_{p_{\sigma}\mu\nu}
+U^{+}_{p_{\sigma}\mu\nu})=\sum\limits_{\mu\nu}\frac{\beta_{\mu\nu}}{2N}\sum\limits_{\{p_{\sigma}\mu\nu\}}
 Tr(U_{p_{\sigma}\mu\nu}
+U^{+}_{p_{\sigma}\mu\nu}),
 \end{equation}
 where $U_{p_{\sigma}\mu\nu}$ is an ordered product of $U_l \in$SU(3)
 around a plaquette, $\mu$ and $\nu$ are the space-time directions,
 $N=3$ is the rank of SU(3). Since the key point of the finite temperature
 lattice gauge field theory 
 is to establish an appropriate partition function with a periodic boundary
 condition of the field in the time-like direction(s), the time-like link(s)
 and space-like links must be treated differently, in the non-isotropic 
 SU(3) lattice gauge model at finite temperature, we suppose that there is 
 only a one-dimension time-like direction which is defined along the $d$th
 direction in the $D$-dimensional non-isotropic space-time lattice. 
 Furthermore, we suppose there is a  $d$-dimensional subspace which 
 are $d$ dimensional space-time
 including the time-like direction in the $D$-dimensional space-time lattice. 
 In the $d$ dimensional subspace the coupling coefficients 
 are assumed identical and called $\beta$, the remaining $(D-d)$ coupling 
 coefficients are also taken to be identical and we call them $\gamma$.
 The reason we choose two different coupling constants $\beta$ and $\gamma$ are
that in one side our familiar universe is an isotropic system, and on the other side
we mainly attempt to show the differences between the non-isotropic gauge system and
our familiar isotropic gauge system. Having these supposition, the Wilson action of the
non-isotropic SU(3) lattice gauge model at finite temperature can be written as: 
\begin{center}
$$S=\frac{\beta}{2N}\left[\sum\limits_{\{p_{\sigma}\mu\nu\}}Tr(U_{p_\sigma}
+U^{+}_{p_\sigma})+\sum\limits_{\{p_{\tau}\mu d\}}Tr(U_{p_\tau}
+U^{+}_{p_\tau})\right]$$
\begin{equation}
+\frac{\gamma}{2N}\left[\sum\limits_{\{p_{\sigma'}\mu'\nu'\}}Tr(U_{p_\sigma'}
+U^{+}_{p_\sigma'})+\sum\limits_{\{p_{\tau'}\mu' d\}}Tr(U_{p_\tau'}
+U^{+}_{p_\tau'})\right],
\end{equation}
\end{center}
here one may consider the first term as an "inside layer" and the second term represents
an interaction between different layers. 
Where  $\mu$, $\nu$($\mu<\nu=1,2,\cdots,d-1$), $\mu'$($\mu'=1,2,\cdots,d-1,d+1,\cdots,D$)
and $\nu'$($\nu'=d+1,\cdots,D$) are space-like directions, ${p_\sigma}$ and ${p_{\sigma'}}$ 
are the space-like plaquettes, and $p_\tau$, $p_{\tau'}$ are the time-like plaquettes.
The summations $\{ \cdots \}$ are taken over all such plaquettes. 

The corresponding partition function of the system is: 
\begin{equation}
Z=\int[dU]e^{S}.
\end{equation}
In general, it is very difficult to calculate the partition function of a certain system.
According to the variational-cumulant expansion method, if one introduce a trial action,
the calculation of partition function will become simple. We introduce a trial action:
$$S_0=\sum\limits_{\sigma}Tr(J^{+}_{\sigma}U_{\sigma}
+J_{\sigma}U^{+}_{\sigma})+\sum\limits_{\sigma'}Tr(J^{+}_{\sigma'}U_{\sigma'}
+J_{\sigma'}U^{+}_{\sigma'})$$
\begin{equation}
+\sum\limits_{\tau}Tr(J^{+}_{\tau}U_{\tau}
+J_{\tau}U^{+}_{\tau}),
\end{equation}
$J_\sigma$, $J_\tau$ and $J_{\sigma'}$  are real variational parameters corresponding
to the space-like link variable $U^{+}_{\sigma}$, the time-like variable
$U^{+}_{\tau}$ in the $d$-dimensional subspace, and the space-like variable
$U^{+}_{\sigma'}$ in the remaining (D-d) directions, respectively. For the 
convenience of calculation, we let $J_\sigma=J_\tau=J$
and $(J_\sigma)_{ij}=J_{ij}=J\delta_{ij}$ and $(J_{\sigma'})_{ij}=J_{\sigma'} \delta_{ij}$.
The corresponding trial partition function of the virtual system
 can be calculated by VCE\cite{qi}:
\begin{equation}
Z(J,J_\sigma')_0=\int[dU]e^{S_0}
\end{equation}
Then the partition function of a genuine system can be achieved via the
trial system as:
\begin{equation}
Z=\int[dU]e^{S}=\int[dU]e^{S_0}
e^{S-S_0}=Z(J_\sigma, J_{\sigma'})_0\left\langle e^{S-S_0} \right\rangle_0,
\end{equation}
where 
$$\langle \cdots\rangle_0=\frac{1}{Z(J,J_\sigma')_0}\int[dU](\cdots)e^{S_0}$$
expresses statistics average in the trial system. 

Now, let us introduce the order parameter, Polyakov line, which can be used as
a criterion of phase transition of a given system:
$$
\langle L\rangle\equiv\left\langle\frac{1}{N}Tr\left(\sum\limits^{N_\tau}_{i=1}
U(x,\tau_i)\right)\right\rangle=\frac{1}{Z}\int[dU]Le^{S}$$
\begin{equation}=\int[dU]L e^{S-S_0}e^{S_0}=\sum\limits_{n=0}^{\infty}\frac{1}{n!}
\langle L(S-S_0)^n\rangle_c\equiv
\sum\limits_{n=0}^{\infty}\frac{1}{n!}\langle L_n\rangle_c
\end{equation}
where $N_\tau=2$ is the periodic condition, 
$\langle \cdots\rangle_c$ indicates the cumulant average in the trial system
which can be expanded in the statistical averages
and the lower order cumulant averages\cite{tan}. For example
$$\langle L_0\rangle_c\equiv\langle L\rangle_0,$$
$$\langle L_1\rangle_c\equiv\langle L(S-S_0)\rangle_c=\langle L(S-S_0)\rangle_0
-\langle L\rangle_0\langle S-S_0\rangle_0,$$
$$
\langle L_2\rangle_c\equiv\langle L(S-S_0)^2\rangle_c=\langle L(S-S_0)^2\rangle_0-
2\langle S-S_0\rangle_0\langle L(S-S_0)\rangle_c-
\langle L\rangle_0\langle (S-S_0)^2\rangle_0,$$
\begin{equation}
\cdots.
\end{equation}
In this work, we calculate the order parameter Polyakov line to 3rd order(n=2). 

\section{Variational treatment}

In principle, we can calculate the Polyakov line to any desired orders using the VCE
method and obtain the exact Polyakov line of the real system without any dependence on 
the variational parameters $J_\sigma$ and $J_{\sigma'}$. In practice, we can only
expand $\langle L\rangle$ to some limited terms and 
the expansion Eq.(8) must be truncated at some order
$n$(in our work, the $\langle L\rangle$ is expanded to $n=2$). 
Thus, the $\langle L\rangle$
turns out to depend on $J_\sigma$ and $J_{\sigma'}$. Then the choice of 
the parameter $J_\sigma$ and $J_{\sigma'}$ will affect the 
rate of convergence of the expansion. We use the convexity inequality to 
obtain the best values of $J_\sigma$ and $J_{\sigma'}$.
First, we introduce the equation of free energy, 
$$Z=Z(J_\sigma, J_{\sigma'})_0\left\langle e^{S-S_0} \right\rangle_0
\equiv e^{-W}=e^{-W_0}\left\langle e^{S-S_0} \right\rangle_0$$
where the $W$ and $W_0=-ln[Z(J_\sigma, J_{\sigma'})_0]$ represent 
the free energy of the real system and trial system.
Then we get the standard convexity inequality equation is 
$$W=-ln[Z]=-ln[Z(J_\sigma,J_{\sigma'})_0\left\langle e^{S-S_0} \right\rangle_0]
\leq-ln[Z(J_\sigma,J_{\sigma'})_0]-\langle S-S_0\rangle_0=W_{eff},$$
where $W_{eff}$ is the upper bound of the free energy.
 
So, the variational parameters $J_\sigma$ and $J_\sigma'$ are determined by 
minimizing the $W_{eff}$:
\begin{equation}
\frac{\delta}{\delta J_{\sigma}}W_{eff}=0,  \frac{\delta}{\delta J_{\sigma'}}W_{eff}=0,
\end{equation}
which lead to
$$-\frac{\beta}{3}(d-1)\displaystyle\left(\frac{Z'_{0}(\sigma)}{6Z_{0}(\sigma)}\right)^3-
\frac{\gamma}{3}(D-d)\left(\frac{Z'_{0}(\sigma)}{6Z_{0}(\sigma)}\right)
\left(\frac{Z'_{0}(\sigma')}{6Z_{0}(\sigma')}\right)^2
+Z_{0}(\sigma)=0,$$
\begin{equation}
-\frac{\gamma}{3}\left[d\left(\frac{Z'_{0}(\sigma')}{6Z_{0}(\sigma)}\right)^2
\left(\frac{Z'_{0}(\sigma')}{6Z_{0}(\sigma')}\right)+
(D-d-1)\left(\frac{Z'_{0}(\sigma')}{6Z_{0}(\sigma')}\right)^3+Z_{0}({\sigma'})=0\right],
\end{equation}
where $Z_{0}(\sigma)$ and $Z_{0}(\sigma')$ are the single link group integral, for
example, the definition of $Z_{0}(\sigma)$ is
$$Z_{0}(\sigma)\equiv Z_{0}(J_{\sigma})=
\int [dU]e^{Tr(J_{\sigma}U^{+}+UJ_{\sigma}^{+})},$$ 
$Z'_{0}(\sigma)$ and $Z'_{0}(\sigma')$ are the corresponding differential of $Z_{0}(\sigma)$
 and $Z_{0}(\sigma')$.
 \section{Result}

Eq.(10) have solutions in the following forms:\\
%\begin{}
$(1)$ $J_\sigma=0$ $(Z_0(J_\sigma)=0, Z'_{0}(J_\sigma)=0)$,
$J_{\sigma'}=0$ $(Z_0(J_{\sigma'})=0, Z'_{0}(J_{\sigma'})=0)$,\\
$(2)$ $J_\sigma>0$ $(Z_0(J_\sigma)>0, Z'_{0}(J_\sigma)>0)$,
$J_{\sigma'}=0$ $(Z_0(J_{\sigma'})=0, Z'_{0}(J_{\sigma'})=0)$,\\
$(3)$ $J_\sigma>0$ $(Z_0(J_\sigma)>0, Z'_{0}(J_\sigma)>0)$,
$J_{\sigma'}>0$ $(Z_0(J_{\sigma'})>0, Z'_{0}(J_{\sigma'})>0)$.\\
Corresponding to the three kinds of solutions there are three kinds of phases:\\
$(a)$ The high-temperature phase corresponding to solution $(1)$. The order parameter is:
$$\langle L\rangle=0.$$
$(b)$ The layer phase corresponding to solution $(2)$. 
The order parameter is:
$$\langle L\rangle=\langle L(\beta)\rangle.$$
$(c)$ The low temperature solution corresponding to solution $(3)$. 
The order parameter is:
$$\langle L\rangle=\langle L(\beta, \gamma)\rangle.$$
Here $\langle L\rangle$ is calculated to the terms of $\beta^2$ and $\gamma^2$ and the
calculation is very tedious, we will not show it here.

Base on the VCE approach to 3rd order, we obtained the layer phase diagrams of the 
non-isotropic SU(3) lattice gauge model at finite temperatures shown 
in Fig.1 and Fig.2. As indicated
by the lattice gauge field theory of zero temperature\cite{fu3}, 
the layer phase of $d=D-1(D\leq 5)$ is unstable, but stable when $d=5(D=6)$ 
 at zero temperature. While in our calculation of finite temperature
 the layer phases exist and 
are stable when $d=D-1(D=4,5)$ for a fixed $N_\tau$ value (here $N_\tau=2$), 
which corresponds to a not very
low temperature, as $N_\tau\longrightarrow \infty$ for zero temperature. It implies that
there must be a transition temperature at which the layer phase turns to be unstable.
Unfortunately, for the case of larger $N_\tau$ value, the calculations with VCE 
are too complicated and difficult to carry out. The triple points of phase diagram
in Fig.1 and Fig.2 are 
located at $\beta_c=9.2,\gamma_c=3.9$ and $\beta_c=6.2,\gamma_c=3.1$, respectively. 

We do not calculate the cases 
when $D\geq 6$, but by our experience, the layer phase will exist in those cases.
For conclusion, using VCE approach, we get information about the layer phase: the layer phase exist
in the non-isotropic non-Abelian SU(3) gauge model, which conform
the former imagination\cite{liang} that the layer phase not only exist
in the non-isotropic U(1) and SU(2) gauge model, but also in other non-isotropic
gauge models. But in this work we can only obtain that the layer phase exists
at a very high temperature in the non-isotropic SU(3) gauge model, we can not 
give the exact temperature at which there is phase transition and
 the layer phase disappear. So, further 
investigation is needed.

\begin{figure}%\begin{center}
   \epsfig{file=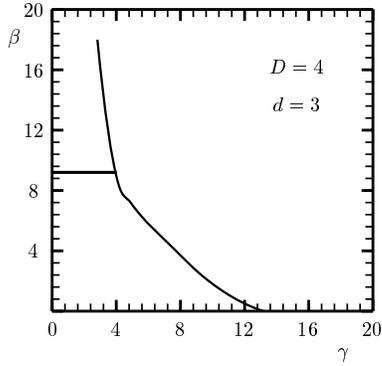, bbllx=130pt,bblly=500pt,bburx=330pt,bbury=700pt,
width=5cm,angle=0}
\caption{\small Phase diagram for the non-isotropic SU(3) gauge model when $D=4$ 
and $d=3$. The upper-right area shows the low-temperature phase, the upper-left area
shows the layer phase and the lower area shows the high-temperature phase.}
%\label{fig}
%\end{center}
\end{figure}
\begin{figure}%\begin{center}
   \epsfig{file=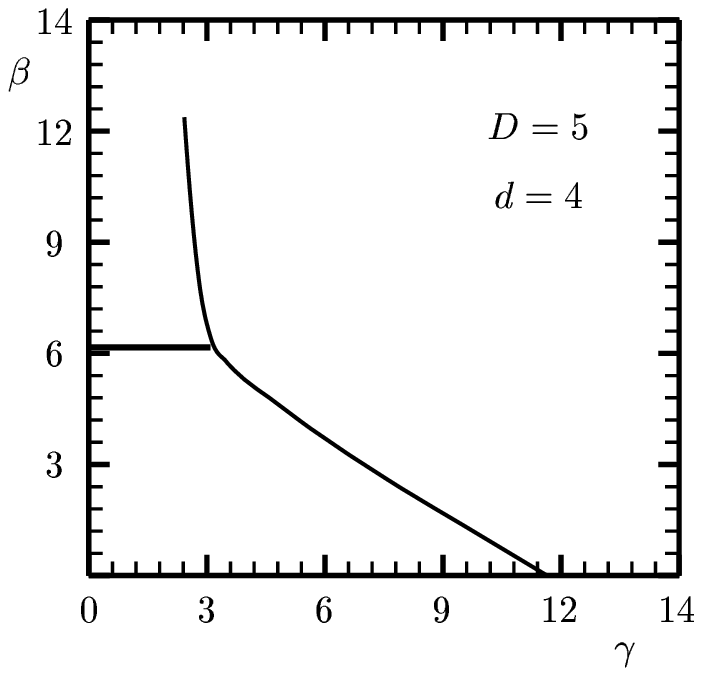, bbllx=130pt,bblly=497pt,bburx=330pt,bbury=700pt,
width=5cm,angle=0}
\caption{\small Phase diagram for the non-isotropic SU(3) gauge model when $D=5$ 
and $d=4$. The upper-right area shows the low-temperature phase, the upper-left area
shows the layer phase and the lower area shows the high-temperature phase.}
%\label{fig}
%\end{center}
\end{figure}
\end{document}